
\magnification = 1200
\def\lapp{\hbox{$ {
\lower.40ex\hbox{$<$}
\atop \raise.20ex\hbox{$\sim$}
}
$}  }
\def\rapp{\hbox{$ {
\lower.40ex\hbox{$>$}
\atop \raise.20ex\hbox{$\sim$}
}
$}  }

\def\krig#1{\vbox{\ialign{\hfil##\hfil\crcr
$\raise0.3pt\hbox{$\scriptstyle \circ$}$\crcr\noalign
{\kern-0.02pt\nointerlineskip}
$\displaystyle{#1}$\crcr}}}
\def\upar#1{\vbox{\ialign{\hfil##\hfil\crcr
$\raise0.3pt\hbox{$\scriptstyle \leftrightarrow$}$\crcr\noalign
{\kern-0.02pt\nointerlineskip}
$\displaystyle{#1}$\crcr}}}
\def\ular#1{\vbox{\ialign{\hfil##\hfil\crcr
$\raise0.3pt\hbox{$\scriptstyle \leftarrow$}$\crcr\noalign
{\kern-0.02pt\nointerlineskip}
$\displaystyle{#1}$\crcr}}}

\def\svec#1{\skew{-2}\vec#1}
\def\Tr{\,{\rm Tr }\,}

\def\g5{\gamma_5}

\def\lp1{{\cal L}_{\pi N}^{(1)}}
\def\lp2{{\cal L}_{\pi N}^{(2)}}
\def\lp3{{\cal L}_{\pi N}^{(3)}}

\topskip=0.60truein
\leftskip=0.18truein
\vsize=8.8truein
\hsize=6.5truein
\tolerance 10000
\hfuzz=20pt

\baselineskip 12pt plus 1pt minus 1pt
\pageno=0
\centerline{\bf CHIRAL CORRECTIONS TO THE S--WAVE}
\smallskip
\centerline{{\bf PION--NUCLEON SCATTERING LENGTHS
 }\footnote{*}{Work supported in part by Deutsche Forschungsgemeinschaft
and by Schweizerischer Nationalfonds.\smallskip}}
\vskip 24pt
\centerline{V\'{e}ronique Bernard}
\vskip 4pt
\centerline{\it Centre de Recherches Nucl\'{e}aires et Universit\'{e}
Louis Pasteur de Strasbourg}
\centerline{\it Physique Th\'{e}orique, Bat. 40A,
BP 20, 67037 Strasbourg Cedex 2,
France}
\vskip 12pt
\centerline{Norbert Kaiser}
\vskip 4pt
\centerline{\it Physik Department T30,
Technische Universit\"at M\"unchen}
\centerline{\it
James
Franck Stra{\ss}e,
W-8046 Garching, Germany}
\vskip 12pt
\centerline{Ulf-G. Mei{\ss}ner\footnote{$^\dagger$}{Heisenberg
Fellow.}}
\vskip 4pt
\centerline{\it Universit\"at Bern,
Institut f\"ur Theoretische Physik}
\centerline{\it Sidlerstr. 5, CH--3012 Bern,\ \ Switzerland}
\vskip 0.5in
\centerline{\bf ABSTRACT}
\medskip
\noindent We calculate the chiral corrections to Weinberg's prediction for the
S--wave $\pi N$ scattering lengths up--to--and--including order ${\cal
O}(M_\pi^3)$ making use of heavy baryon chiral perturbation theory. For the
isospin--odd scattering length $a^-$ these corrections are small and bring the
prediction closer to the empirical value. In the case of the isospin--even
$a^+$ large cancellations appear so that the ${\cal O}(M_\pi^3)$ result depends
sensitively on certain resonance parameters which enter the calculation of the
contact terms present at next--to--leading order.
\medskip
\vfill
\noindent BUTP--93/09\hfill April 1993

\noindent CRN 93--13
\eject
\baselineskip 14pt plus 1pt minus 1pt
\noindent 1. One of the most spectacular successes of current algebra in the
sixties was Weinberg's prediction [1] for the S--wave pion--nucleon scattering
lengths, $a_{1/2} = M_\pi / 4 \pi F_\pi^2 = -2 a_{3/2} = 0.175 \, M_\pi^{-1}$,
with $M_\pi$ the
physical pion mass and $F_\pi$ the pion decay constant. Tomozawa [2] also
derived the sum rule $a_{1/2} - a_{3/2} = 3 M_\pi / 8 \pi F_\pi^2 = 0.263$
$M_\pi^{-1}$. Empirically, the combination $(2 a_{1/2} + a_{3/2}) / 3$ is best
determined. The Karlsruhe--Helsinki group gives  $0.083 \pm 0.004$ $M_\pi^{-1}$
[3] consistent with the pionic atom measurement [4] of $0.086 \pm 0.004$
$M_\pi^{-1}$. The value of $a_{1/2} - a_{3/2}$ is more uncertain. The KH
analysis leads to $0.274 \pm 0.005$ $M_\pi^{-1}$ [5] whereas the VPI group has
recently given a larger value [6]. Their analysis was critically reexamined by
H\"ohler [7]. In what follows, we will use the central values from the work of
Koch [3], namely $a_{1/2} = 0.175$ $M_\pi^{-1}$ and $a_{3/2} = -0.100$
$M_\pi^{-1}$. The agreement of the current algebra predictions with these
numbers is rather spectacular. However, in the last decade it has become clear
that current algebra is only the first term in a systematic expansion of the
QCD Green functions in powers of (small) external momenta and the (light) quark
masses [8,9]. Therefore, one would like to know what the next--to--leading
order corrections to the original predictions are. This is exactly the question
we will address here. The basic framework to perform the calculation of these
corrections is baryon chiral perturbation theory which makes use of an
effective
Lagrangian of the asymptotically observed fields. In particular, we use the
heavy fermion formulation [10] in which the nucleons are considered as static
sources and one has a one--to--one correspondence between the loop and the
small momentum expansion. We will work out one loop and counter term
contributions up--to--and--including order $M_\pi^3$ based on the power
counting scheme developed by Weinberg [8] and later extended to the baryon
sector by Gasser et al. [11]. For a review on these methods, the reader is
referred to ref.[12].
\bigskip
\noindent 2. Consider the $\pi N$ forward scattering amplitude for a nucleon at
rest with its four--velocity given by $v_\mu = (1,0,0,0)$.\footnote{*}{Remember
that we treat the nucleons as very heavy fields.} Denoting by $b$ and $a$ the
isospin of the outgoing and incoming pion, in order, the scattering amplitude
takes the form
$$T^{ba} = T^+ (\omega) \delta^{ba} + T^- (\omega) i \epsilon^{bac} \tau^c
\eqno(1)$$
with $q$ the pion four--momentum and $\omega = v \cdot q = q_0$. Under
crossing $(a \leftrightarrow b, \, q \to -q )$ the functions $T^+$ and $T^-$
are even and odd, respectively, $T^\pm (\omega) = \pm T^\pm (-\omega)$. At
threshold one has $\svec q = 0$ and the pertinent scattering lengths are
defined by
$$a^\pm = {1 \over 4 \pi} \bigl( 1 + {M_\pi \over m} \bigr)^{-1}\, T^\pm (
M_\pi ) \eqno(2)$$
with $m$ the nucleon mass. The S--wave scattering lengths for the total $\pi N$
isospin 1/2 and 3/2 are related to $a^\pm$ via
$$a_{1/2} = a^+ + 2 a^- , \quad a_{3/2} = a^+ -  a^-  \eqno(3)$$
The abovementioned central empirical values translate into  $a^+ = -0.83 \cdot
10^{-2} \, M_\pi^{-1}$ and $a^+ = 9.17 \cdot 10^{-2} \, M_\pi^{-1}$. In what
follows, we will not exhibit the canonical units of $10^{-2} \, M_\pi^{-1}$.
The benchmark values are therefore $a^+ = -0.83 \pm 0.38$ and $a^- = 9.17 \pm
0.17$\footnote{*}{We have added the errors obtained for $a_{1/2}$ and $a_{3/2}$
in quadrature.} compared to
the current algebra predictions of $a^+ = 0$ and $a^- = 8.76$ (using $M_\pi =
138$ MeV and $F_\pi = 93$ MeV).
\bigskip
\noindent 3. To calculate the scattering lengths, we use the effective
pion--nucleon Lagrangian. We work in flavor SU(2) and in the isospin limit
$m_u = m_d = \hat m$. The pion fields are collected in the matrix $U(x) =
\exp[i{\vec \tau}\cdot{\vec \pi}(x) / F_\pi] = u^2 (x)$. The effective
Lagrangian to order ${\cal
O}(q^3)$, where $q$ denotes a genuine small momentum or a quark mass, reads
$$\eqalign{{\cal L}_{\pi N} &= {\cal L}_{\pi N}^{(1)} +
{\cal L}_{\pi N}^{(2)} + {\cal L}_{\pi N}^{(3)} \cr
{\cal L}_{\pi N}^{(1)} &= {\bar H} ( i v\cdot D + g_A S \cdot u ) H \cr
{\cal L}_{\pi N}^{(2)} &= c_1\, {\bar H}H \Tr (\chi_+) + \bigl( c_2 - {g_A^2
\over 8 m} \bigr) {\bar H} v\cdot u\, v\cdot u H + c_3\, {\bar H} u\cdot u H
\cr {\cal L}_{\pi N}^{(3)} &= \bigl( b_1 - {g_A^2\over 32 m^2}\bigr) \, v^\mu
v^\lambda
v^\rho {\bar H} K_{\mu \lambda \rho} H + b_2 \, v^\rho {\bar H} K^\mu_{\mu
\rho} H + b_3 \, v^\rho {\bar H} K^\mu_{\rho \mu} H \cr} \eqno(4)$$
with
$$\eqalign{u_\mu &= i u^\dagger \nabla_\mu U u^\dagger \cr
u_{\mu \lambda} &= i u^\dagger \nabla_\mu \nabla_\lambda U u^\dagger \cr
K_{\mu \lambda \rho} &= i [ u_{\mu \lambda} , u_\rho ] \cr} \eqno(5)$$
where $H$ denotes the heavy nucleon field, $S_\mu$ the covariant spin--operator
subject to the constraint $v \cdot S = 0$, $\nabla_\mu$ the covariant
derivative acting on the pions and we adhere to the notations of ref.[13]. The
superscripts (1,2,3) denote the chiral dimension. The lowest order effective
Lagrangian is of order ${\cal O}(q)$. The one loop contribution is suppressed
with respect to the tree level by $q^2$ thus contributing at ${\cal O}(q^3)$.
In addition, there are contact terms of order $q^2$ and $q^3$ with coefficients
not fixed by chiral symmetry. Notice furthermore that one in addition has to go
further in the $1/m$ expansion of the relativistic tree level graphs as can be
seen from
the terms which come together with the ones proportional to $c_2$ and $b_1$. At
next--to--leading order, all these
terms have to be retained. Due to crossing symmetry, ${\cal L}_{\pi N}^{(2)}$
contributes only to $T^+(\omega)$ whereas ${\cal L}_{\pi N}^{(3)}$ solely
enters $T^-(\omega)$.
We have not exhibited the standard meson Lagrangian ${\cal L}_{\pi \pi}^{(2)} +
{\cal L}_{\pi \pi}^{(4)}$. The contact terms appearing in ${\cal L}_{\pi
\pi}^{(4)}$ only contribute to the shift of the pion mass, the pseudoscalar
coupling $G_\pi$ and the pion decay constant from their chiral limit to the
physical values. For the following, we define
$$ L = {M_\pi \over 8 \pi F_\pi^2}\, , \quad \mu = {M_\pi \over m} \eqno(6)$$
The calculation of the scattering lengths is straightforward. For the
isospin--odd $a^-$ one arrives at
$$\eqalign{a^- &= a^- (M_\pi) + a^- (M_\pi^2) + a^- (M_\pi^3) \cr
a^- (M_\pi) &= L \, , \quad a^- (M_\pi^2)  = - L \mu  \cr
a^- (M_\pi^3) &= L \mu^2 \bigl[1 + {g_A^2 \over 4}\bigr] + {L^2 M_\pi \over
\pi} \bigl[ 1 - 2 \ln { M_\pi \over \lambda}\bigr] - 64 \pi L^2 M_\pi F_\pi^2
\bigl[
b_1^r (\lambda) + b_2 + b_3 \bigr] \cr} \eqno(7)$$
with $\lambda$ the scale of dimensional regularization. While the constants
$b_2$ and $b_3$ are finite, $b_1$ has to be renormalized as follows to render
the isospin--odd scattering amplitude $T^- (\omega)$ finite,
$$\eqalign{b_1 &= b_1^r (\lambda) - { K \over 2 F^2} \cr
K &= {\lambda^{d-4} \over 16 \pi^2} \biggl[ {1 \over d-4} + {1\over 2} \bigl(
\gamma_E - 1 -\ln 4 \pi \bigr) \biggr] \cr} \eqno(8)$$
Here, $F$ is the pion decay constant in the chiral limit. In what follows, we
will use $\lambda = m_\Delta = 1.232$ GeV, motivated by the resonance
saturation principle. Of course, physical observables
do not depend on this particular choice. Notice that the contact term
contributions are suppressed by a factor $M_\pi^2$ with respect to the leading
current algebra term. Matters are different for the  isospin--even scattering
length $a^+$. It consists of contributions of order $M_\pi^2$ and $M_\pi^3$,
$$\eqalign{a^+ &= a^+ (M_\pi^2)+  a^+ (M_\pi^3) \cr
a^+ (M_\pi^2) &= 32 \pi F_\pi^2 L^2 \bigl( c_2 + c_3 - 2c_1 - {g_A^2 \over 8m}
\bigr) \cr
a^+ (M_\pi^3) &= {3 \over 4} g_A^2 L^2 M_\pi   - 32 \pi F_\pi^2 L^2  \mu
\bigl( c_2 + c_3 - 2c_1 - {g_A^2 \over 8m}\bigr) \cr}\eqno(9)$$
The coefficients $c_{1,2,3}$ are all finite. From the form of eq.(9) it is
obvious that the contact terms play a more important role in the determination
of $a^+$ than for $a^-$.

\noindent 4. The most difficult task is to pin down the various low-energy
constants appearing in eqs.(7) and (9). Let us first consider $c_{1,2,3}$. The
coefficient $c_1$ can be unambigously fixed from the pion--nucleon
$\sigma$--term [13],
$$c_1 = -{1 \over 4 M_\pi^2} \biggl[ \sigma_{\pi N}(0) + {9 g_A^2 M_\pi^3 \over
64 \pi F_\pi^2}\biggr] = -0.87 \pm 0.11  \,{\rm GeV}^{-1} \eqno(10)$$
using $g_A = 1.26$ and $\sigma_{\pi N}(0) = 45 \pm 8$ MeV [14]. To estimate the
remaining constants, we make use of the principle of resonance saturation [15].
It states that to a high degree of accuracy the low--energy constants can be
calculated from resonance exchanges by integrating out the heavy fields from an
effective Lagrangian of the pions chirally coupled to the various resonances.
In the meson sector, this has been shown to work very well. We extend this
method to the baryon sector since it is essentially the only method of
estimating the unknown coefficients. From the meson sector, scalar meson
exchange can contribute to $c_1$ and $c_3$,
$$c_1 -{1 \over 2} c_3 \big|_S = c_1 - c_1 {c_d \over c_m} \eqno(11)$$
with $2c_d / c_m = L_5 / L_8$ [15]. The central values for the parameters $c_d$
and $c_m$ given in ref.[15] are $c_m = 42$ MeV and $c_d = 32$ MeV, i.e.
$2c_d / c_m = 1.56$. However, within the uncertainty of $L_5$ and $L_8$, this
ratio can vary between 0.75 and 2.25. In addition, intermediate $\Delta(1232)$
states give a contribution to $c_2 + c_3$. The general $\pi \Delta N$--vertex
can be written as
$${\cal L}_{\pi \Delta N} = {g_{\pi \Delta N} \over 2 m} \bar{\Delta}^\mu_a
\bigl[ g_{\mu \nu} - ( Z + {1\over 2}) \gamma_\mu \gamma_\nu \bigr]
\partial^\nu \pi^a \, N + \, {\rm h.c.} \eqno(12)$$
where $\Delta_\mu$ denotes the Rarita--Schwinger field and $Z$ parametrizes the
off--shell behaviour of the spin--3/2 field.\footnote{$^\star$}{It is mandatory
to use here the relativistic formulation of the spin--3/2 field since otherwise
one would miss a contribution of order $M_\pi^2$ to $a^+$.} While Peccei [16]
fixed $Z = -1/4$, a more recent phenomenological analysis of Benmerrouche et
al.[17] gives a rather wide range, $-0.8 \le Z \le 0.3$. Using the empirically
well--fulfilled SU(4) coupling constant relation $g_{\pi \Delta N} = 3 g_{\pi
N} / \sqrt{2}$ together with the Goldberger--Treiman relation, we can cast the
$\Delta (1232)$ contribution to $c_2 + c_3$ into the form
$$c_2 + c_3 \big|_\Delta = -{g_A^2 \over 2 m_\Delta^2} ({1\over 2} - Z) \biggl[
2 m_\Delta (1 + Z) + m ( {1 \over 2} - Z) \biggr]\eqno(13)$$
Clearly, this dependence on $Z$ is one of the major sources of uncertainty in
fixing the values of the contact terms $c_2$ and $c_3$. Furthermore, there is
also a contribution from the Roper $N^* (1440)$ resonance to $c_2 +c_3$,
$$c_2 + c_3 \big|_{N^*} = -{g_A^2 R \over 16 (m+m^*)} \eqno(14)$$
with $R = 1 \ldots 1/4$ for $g_{\pi N N^*} = (1/2 \ldots 1/4) g_{\pi N}$ [18].
In complete analogy, one also has a $\Delta$ and $N^*$ contribution to the
low--energy constants $b_1^r$, $b_2$ and $b_3$. At $\lambda = m_\Delta$, it can
be written as
$$b_1^r (\lambda) + b_2 + b_3 \big|_{\lambda = m_\Delta} = -g_A^2 \biggl[{( Z -
{1 \over 2})^2
\over 8 m^2_\Delta} + {R \over 32 (m + m^* )^2} \biggr]
\eqno(15)$$
Finally, other baryon resonances have been neglected since their couplings are
either very small or poorly (not) known.\footnote{*}{A remark on the
$\rho$--meson is in order. The chiral power counting enforces a $\rho \pi \pi$
vertex of order $q^2$ of the form ${\cal L}^{(2)}_{\rho \pi \pi} =
g_{\rho \pi \pi} \Tr ( \rho_{\mu \nu} [ u^\mu , u^\nu ])$ [15]. In forward
direction the contraction of the $\rho$--meson propagator with the
corresponding $\rho \pi \pi$ matrix element vanishes. Therefore, one has no
explicit $\rho$--meson induced contributions to $a^-$ of order $q^2$ and
$q^3$.}
It is interesting to note that
for $Z = -1/4$ and $R=1$, the $N^* (1440)$ contribution to $c_2+c_3$ and to
$b_1^r (\lambda) + b_2 + b_3$ is 4 and 12 per cent of the
$\Delta$--contribution,
respectively. However, in the latter case the contact terms play a much less
pronounced role as already discussed.
\bigskip
\noindent 5. We now present our numerical results. Consider first the
scattering length $a^-$. Using $M_\pi = 138$ MeV, $F_\pi = 93$ MeV, $m = 938.9$
MeV, $Z = -1/4 $ and $R = 1$, we find
$$a^- = (8.76 - 1.29 + 1.69) \cdot 10^{-2}\,M_\pi^{-1}
= 9.16 \cdot 10^{-2}\,M_\pi^{-1} \eqno(16)$$
where we have explicitely shown the contributions of order $M_\pi$,
$M_\pi^2$ and $M_\pi^3$ (and momentarily reinstated the units). The total
result is in good agreement with the
empirical value. The largest part of the $M_\pi^3$   term comes from the pion
loop diagrams, it amounts to 1.31 for $\lambda = m_\Delta$. Varying $Z$ within
its band of allowed values, $a^-$ varies by $\pm 0.15$. The dependence on $R$
shows up only in the third digit and is thus irrelevant. The one--loop
corrections bring the the lowest order value closer to the empirical one.
One might argue that since the $M_\pi^3$ contribution is even larger than the
$M_\pi^2$ one, the chiral series has no chance of converging.
However, the important one loop effect can only show up at order $M_\pi^3$
due to the chiral counting. The two--loop contribution carries an explicit
factor $M_\pi^5$ and is therefore expected to be much smaller. Clearly, one
would like to perform a calculation beyond ${\cal O}(q^3)$ as done here, but
this is beyond the scope of this paper. In the case of the isospin--even
scattering length $a^+$, the situation is much less satisfactory.  There are
large cancellations between the loop contribution and the $1/m$ suppressed
kinematical terms of order $M_\pi^2$ and $M_\pi^3$. For $M_\pi = 138$ MeV,
these two amount to $0.91 - 0.87 = 0.04$ in the conventional units. Therefore,
the role of the contact terms is even further magnified. In fig.1, we show
$a^+$ as a function of $Z$ for our standard input and $R=1$. The empirical
value of $a^+$ can be obtained for a small and positive value of $Z$, $Z \simeq
0.15$. As shown in fig.2, $a^+$ varies
also strongly as the ratio $2c_d/ c_m$
changes. Setting $Z=0$, the empirical value results for $2c_d/ c_m = 1.32$, not
far from its central value of 1.56. Clearly, a better understanding of these
resonance parameters is necessary before one can draw a final conclusion on the
accuracy of the chiral expansion for $a^+$. It is, however, gratifying that
slight variations of these resonance parameters allow one to obtain the
empirical value.
\bigskip \noindent
6. To summarize, we have used heavy baryon chiral perturbation theory to
calculate the corrections up--to--and--including order $M_\pi^3$ to Weinberg's
lowest order (current algebra) predictions for the two S--wave pion--nucleon
scattering lengths, $a^\pm$. To estimate the strength of the various
contact terms, we have made use of the principle of resonance saturation which
is known to work very accurately in the meson sector. The main results of this
investigation are:
\medskip
\item{$\bullet$}{The chiral corrections to the isospin--odd scattering length
$a^-$ are small and positive and move the lowest order prediction closer to the
empirical value. The main effect comes form the pion loop diagrams.    The
contact term contribution is relatively small, thus masking the uncertainty in
estimating the coefficients which appear together with these terms.}
\medskip
\item{$\bullet$}{The situation is very different for the isospin--even
scattering length $a^+$. The contact term contribution completely dominates the
chiral expansion since there is a large cancellation between the one--loop and
the
kinematical corrections. The total result for $a^+$ is very sensitive to some
of the resonance parameters, the empirical value of $a^+$ can, however, be
obtained by reasonable choices of these.}
\medskip    \noindent
Evidently, further investigations have to go into two directions. First, a
calculation beyond ${\cal O}(q^3)$ has to be performed to find out how fast the
chiral expansion of the scattering lengths converges. Second, a better
understanding of the coefficients of the contact terms appearing at order $q^2$
and higher is necessary to further pin down the prediction for $a^+$. We hope
to come back to these topics in the near future.
\vfill \eject
\noindent
{\bf References}
\bigskip
\item{1.}S. Weinberg, {\it Phys. Rev. Lett.\/} {\bf 17} (1966) 616.
\smallskip
\item{2.}Y. Tomozawa, {\it Nuovo Cim.\/} {\bf 46A} (1966) 707.
\smallskip
\item{3.}R. Koch, {\it Nucl. Phys.\/} {\bf A448} (1986) 707.
\smallskip
\item{4.}W. Beer et al., {\it Phys. Lett.\/} {\bf B261} (1991) 16.
\smallskip
\item{5.}G. H\"ohler, in Land\"olt--B\"ornstein, vol.9 b2, ed. H. Schopper
(Springer, Berlin, 1983).
\smallskip
\item{6.}R.L. Workman, R.A. Arndt and M. Pavan,
{\it Phys. Rev. Lett.\/} {\bf 68} (1992) 1653.
\smallskip
\item{7.}G. H\"ohler, Karlsruhe preprint TTP 92-21, presented at the Workshop
on $\pi N$ Scattering, Blacksburg, Va., August 1992.
\smallskip
\item{8.}S. Weinberg, {\it Physica\/} {\bf 96A} (1979) 327.
\smallskip
\item{9.}J. Gasser and H. Leutwyler, {\it Ann. Phys. (N.Y.)\/}
 {\bf 158} (1984) 142;
 {\it Nucl. Phys.\/}
 {\bf B250} (1985) 465.
\smallskip
\item{10.}E. Jenkins and A.V. Manohar, {\it Phys. Lett.\/} {\bf B255} (1991)
558.
\smallskip
\item{11.}J. Gasser, M.E. Sainio and A. ${\check {\rm S}}$varc,
{\it Nucl. Phys.\/}
{\bf B307} (1988) 779.
\smallskip
\item{12.}
Ulf-G. Mei{\ss}ner, "Recent Developments in Chiral Perturbation Theory", Bern
University preprint BUTP-93/01, 1993, to appear in {\it Rep. Prog. Phys.}
\smallskip
\item{13.}V. Bernard, N. Kaiser, J. Kambor
and Ulf-G. Mei{\ss}ner, {\it Nucl. Phys.\/} {\bf B388} (1992) 315.
\smallskip
\item{14.}J. Gasser, H. Leutwyler and M.E. Sainio, {\it Phys. Lett.\/}
{\bf B253} (1991) 252.
\smallskip
\item{15.}G. Ecker, J. Gasser, A. Pich and E. de Rafael,
{\it Nucl. Phys.\/} {\bf B321} (1989) 311.
\smallskip
\item{16.}R.D. Peccei, {\it Phys. Rev.\/} {\bf
176} (1968) 1812.
\smallskip
\item{17.}M. Benmerrouche, R.M. Davidson and N.C. Mukhopadhyay,
{\it Phys. Rev.\/} {\bf C39} (1989) 2339.
\smallskip
\item{18.}T. Ericson and W. Weise, "Pions and Nuclei", Clarendon Press, Oxford,
1988.
\bigskip
\noindent
\noindent{\bf Figure Captions}
\bigskip
\item{Fig.1}{The scattering length $a^+$ as a function of $Z$ in units of
$10^{-2}\,    M_\pi^{-1}$. The input is specified in the text. The empirical
range is also shown (the solid line gives the central value).}
\medskip
\item{Fig.2}{The scattering length $a^+$ as a function of $2c_d/c_m$ in units
of $10^{-2}\,
M_\pi^{-1}$. The input is specified in the text. For notations, see fig.1.}
\medskip
\vfill \eject \end